\documentclass[twocolumn,showpacs,preprintnumbers,amsmath,amssymb,prb]{revtex4}
\usepackage{graphicx}
\usepackage{amsmath, amsthm, amssymb}
\usepackage{latexsym}
\usepackage{amssymb}
\usepackage{bm}
\usepackage{dcolumn}
\usepackage{epstopdf}
\usepackage{color}

\begin{document}

\title{$^{75}$As NMR of Ba(Fe$_{0.93}$Co$_{0.07}$)$_{2}$As$_{2}$ in High Magnetic Field}

\author{Sangwon Oh$^{1}$, A. M. Mounce$^{1}$, S. Mukhopadhyay$^{1}$, W. P. Halperin$^{1}$, A. B. Vorontsov$^{2}$, S. L. Bud\'{}ko$^{3}$, P. C. Canfield$^{3}$, Y. Furukawa$^{3}$, A. P. Reyes$^{4}$, P. L. Kuhns$^{4}$}

\affiliation{$^1$Department of Physics and Astronomy, Northwestern University, Evanston, Illinois 60208, USA \\
$^2$Department of Physics, Montana State University, Bozeman, Montana 59717, USA\\
$^3$Ames Laboratory US DOE and Department of Physics and Astronomy, Iowa State University, Ames, IA 50011, USA\\
$^4$National High Magnetic Field Laboratory, Tallahassee, Florida 32310, USA}
\date{Version \today}

\begin{abstract}
The superconducting state of an optimally doped single crystal of
Ba(Fe$_{0.93}$Co$_{0.07}$)$_2$As$_2$ was investigated by $^{75}$As NMR in high
magnetic fields from 6.4 T to 28 T. It was found that the Knight shift is least
affected by vortex supercurrents in high magnetic fields, $H>11$ T, revealing

slow, possibly higher order than linear, increase with temperature at   
$T \lesssim 0.5 \, T_c$, with $T_c \approx 23 \, K$. 
This is consistent with the extended $s$-wave state with $A_{1g}$ symmetry but 
the precise details of the gap structure are harder to resolve.  
Measurements of the NMR spin-spin
relaxation time, $T_2$, indicate a strong indirect exchange interaction at all
temperatures.  Below the superconducting transition temperature abrupt changes in vortex dynamics
lead to an anomalous dip in $T_2$  consistent with vortex freezing from which
we obtain the vortex phase diagram up to $H = 28$ T.  
\end{abstract} 

\pacs{ }

\maketitle

\section{Introduction}
As a result of the discovery of pnictide superconductors by 
Kamihara {\it et al.},~\cite{kamihara08}  in 2008 there has been intense interest  in this new
family of materials.  In the past two years their quality has
greatly improved and large single crystals have become available. However, the
nature of the superconducting state and the corresponding gap structure of the
pnictide superconductors is not settled.

While the pairing mechanism, related to magnetic fluctuations, and basic structure 
of the extended $s$-wave state are generally agreed upon, 
\cite{MazinSchmalian,Mazin:2008splus,Chubukov:2008rg,Wang:2009frg}
details of possible gap anisotropy are currently actively investigated, 
often with opposite conclusions. 

Experiments from  angle resolved photoemission spectroscopy
(ARPES),~\cite{ding08,kondo08,terashima09} Andreev-reflection
spectroscopy,~\cite{szabo09} and specific heat~\cite{mu09} measurements support
the fully-gapped model. But penetration depth
measurements,~\cite{gordon09,martin09, fletcher09, hicks09} nuclear quadrupole
resonance (NQR), and specific heat measurements~\cite{fukazawa09} have been
interpreted in terms of nodal gap structure.  These measurements indicate 
that the
superconducting order parameter might be fundamentally different in different
classes of these materials, further complicating interpretation of the
experiments.

Moreover, even in a single compound, but with different amount of doping, 
various gap signatures have been seen, from fully gapped, to nodal character. 
For example, calorimetric,\cite{DJJang:2010_calor,Gofryk:2010_anneal,Gofryk:2010_dopdep} 
transport\cite{Tanatar:2010doping}
and optical conductivity\cite{Fischer:2010_122od} 
measurements 
in overdoped Ba(Fe$_{1-x}$Co$_x$)$_2$As$_2$ ($x>0.1$) show nodal character, 
and similarly in underdoped samples, \cite{LanLuan:2010,Gofryk:2010_anneal} 
whereas at optimal doping $(x\sim 0.06-0.08)$ results 
are contradictorily interpreted either in terms of a more or less isotropic  
gap\cite{kim_can10,LanLuan:2010,Gofryk:2010_anneal} 
or a strongly anisotropic gap.~\cite{mazin:2010_122nodes,Lobo:2010_122optic}

The fact that the gap structure might be doping-dependent is a possibility. 
Theory predicts that in multi-band 
materials, such as pnictides, one may naturally find states with ``accidental'' nodes, 
even in the most symmetric $A_{1g}$ configuration. 
In most cases the nodes appear on the outer electronic Fermi surface sheets, 
depending on the values of the interaction parameters, which are functions of 
doping.~\cite{Kuroki:2008band5,Seo:2008splus,Maier:2009,Chubukov:2009nodes,Thomale:2009nodes}
The outstanding question now is what is the form and position of nodal lines 
on the electron Fermi surface.
\cite{anton10,ChubukovEremin,mazin:2010_122nodes,graser:2010_3d122}
Other nodal locations (e.g. on central hole pocket) have also been considered.~\cite{Graser:2009degenr,Goswami:2009degenr}

In this work we report nuclear magnetic resonance (NMR) measurements on single crystals of Ba(Fe$_{0.93}$Co$_{0.07}$)$_2$As$_2$.  NMR has advantages in that it can probe bulk characteristics, is sensitive to the electronic structure within a penetration depth of the surface of a single crystal, 
$\gtrsim 100$ nm, 
and is less susceptible to surface conditions as might be the case in STM and ARPES experiments. In principle, Knight shift data from NMR can be an appropriate indicator of fully gapped or nodal gap structure. For the former, the Knight shift has the temperature dependence of the Yosida function~\cite{yosida58} or normal fluid density, and for a 
gap with line nodes the temperature dependence should be linear at low temperatures.~\cite{zheng02} Our goal was to investigate the Knight shift at sufficiently high magnetic fields that vortex contributions to the temperature dependence of the local field are minimized, and at sufficiently  low temperatures to identify the gap structure.   It is also important to determine independently that the vortex structures have a static distribution, that is to say there is a solid vortex state.  Then the vortex configuration is stable and cannot introduce temperature dependence to the lineshape, affecting the Knight shift analysis.  Consequently we have measured the spin-spin relaxation time, $T_2$, as a function of temperature and magnetic field and found it to be a good indicator for the onset of vortex dynamics.  We focus on measurements of the Knight shift below the temperature where vortices become frozen.

\section{Experimental Methods}
 We have performed $^{75}$As NMR studies on a single crystal of
Ba(Fe$_{1-x}$Co$_{x}$)$_2$As$_2$ ($x=0.074$) with zero field $T_c= 22.5$ K, varying temperatures from 2 K to 200 K with external 
magnetic field from $H = 6.4$ T to $H = 28$ T parallel to the $c$-axis of the sample. The measurements were performed at Northwestern University and the National High Magnetic Field Laboratory in Tallahassee, Florida. The crystal was grown at Ames Laboratory by the self-flux method,~\cite{ni08} having dimensions of 3.9$\times$5.3$\times$0.7 mm$^{3}$ and mass of 49.3 mg.  A Hahn echo  sequence ($\pi /2-\tau-\pi-\tau-$echo) was used for spectra, Knight shift, linewidth, and some of our spin-spin relaxation time, $T_{2}$, measurements, where $\tau$ is the delay time.  A typical $\pi/2$ pulse length was 4 $\mu$s, defined as that which gave maximum echo intensity. We used frequency sweeps when a single pulse did not cover a sufficiently wide frequency range. Delay times for acquisition of the spectra were varied from 70 to 200 $\mu$s depending on temperature. The Knight shift and linewidth
measurements were determined from gaussian fits to the line shape. The gyromagnetic ratio of the bare nucleus $^{75}$As, $\gamma=$ 7.2919 MHz/T, was used as the reference for the Knight shift. Linewidths were defined as full-width-at-half-maximum (FWHM) of the gaussian distribution. For $T_{2}$ measurements from the Hahn echo sequence, we varied $\tau$  from 100 $\mu$s to 1.40 ms, and determined the rate from the initial portion of the recovery.  In high magnetic fields, $H > 17$ T in a resistive magnet, we used the Carr-Purcell-Meiboom-Gill (CPMG) sequence which is resistant to magnetic field fluctuations,~\cite{sigmund01}  ($\pi_{x}/2-\tau-\pi_{y}-\tau-$echo$-\tau-\pi_{y}-. . . $ and $x,y$ indicate orthogonal RF phases).  Typically, the refocusing time, $2\tau$, was chosen to be 250 $\mu$s. In the course of these spin-spin relaxation measurements our comparison of CPMG and Hahn echo experiments revealed the existence of strong field fluctuations intrinsic to the sample.  In all cases, the pulse sequence repetition time in the superconducting state was taken to be of order the spin-lattice relaxation time,~\cite{imai08} $T_{1}$, and was increased with decreasing temperature.

\section{NMR Spectrum and Knight Shift}
\begin{figure}[!ht]
\centerline{\includegraphics[width=0.35\textwidth]{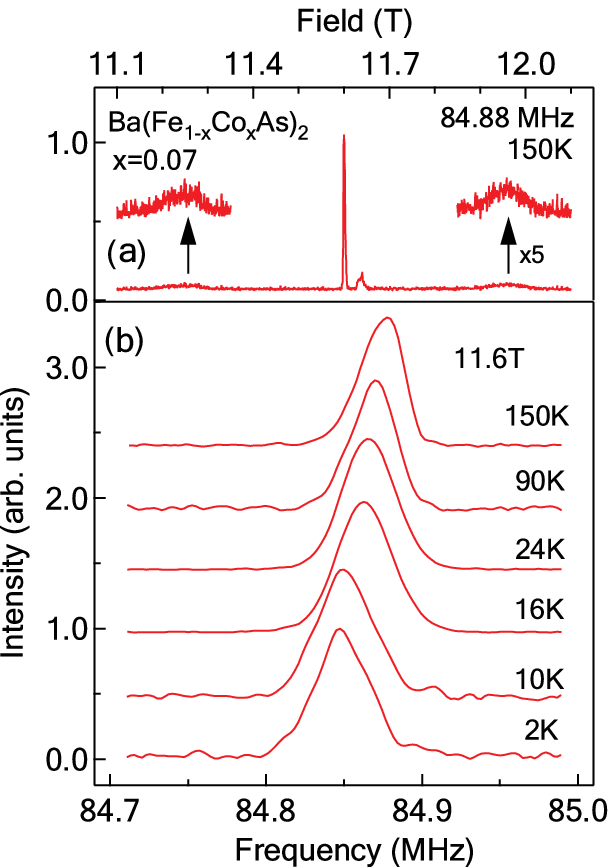}}
\caption {$^{75}$As NMR spectra of Ba(Fe$_{0.93}$Co$_{0.07}$)$_2$As$_2$.  a) Single crystal spectra were obtained 
from a field sweep at 84.88 MHz. From the quadrupolar satellite transitions, shown magnified for clarity, $\nu_{Q}$ was found to be 2.58 MHz 
(0.35 T).  b) Spectra  of the central transition are shown at different temperatures, normalized in area.}
\label{fig1}
\end{figure}

Our field swept spectra of $^{75}$As NMR and the shift of their peak
frequencies with temperature are shown in Fig.~\ref{fig1}.  From the satellite transition we find the quadrupolar coupling, $\nu_{Q} = 2.58$ MHz (0.35 T), similar to the value reported by Ning {\it et al.}~\cite{imai08} The quadrupolar satellite transitions are suppressed because of disorder in the electric field gradient  (EFG) tensor, most likely from Co doping, from which the central transition is largely immune. In part a) of this figure there is a small component of the spectrum having a negligible Knight shift (0.003\%). It is clear that these nuclei are from impurity or defect  sites having a different orbital electronic shift. The spectra in the lower panel are obtained from the central transition.  Similar observations were made by Ning {\it et al.}~\cite{imai08}

\begin{figure}[!ht]
\centerline{\includegraphics[width=0.4\textwidth]{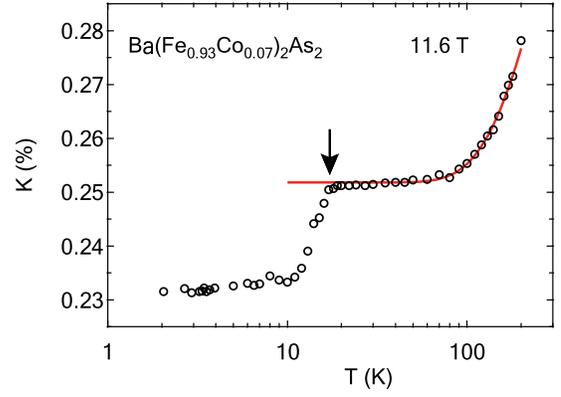}}
\caption {Knight shift of Ba(Fe$_{0.93}$Co$_{0.07}$)$_2$As$_2$.   The temperature dependence of the Knight shift consists of three parts: at low temperature (below the vortex freezing temperature), near the superconducting transition (the arrow indicates $T_c$), and in the normal state.  The drop in Knight shift at the superconducting transition is evidence for a spin-singlet state. In the normal state the Knight shift  follows an activated behavior, with activation energy of 394 K. The solid curve (red) is the fit to this activated process.  The low temperature behavior shows a decreasing Knight shift from which we find $K(0)=0.231\%$ by extrapolating to zero temperature.  }
\label{fig2}
\end{figure}

Our measurements of the temperature variation of the total Knight shift, $K$, of the $^{75}$As central transition  are presented in Fig.~\ref{fig2}.  Spin-singlet pairing is
evident from the sharp decrease of $K(T)$ at $T_c$. The onset of this decrease at 18 K in 11.6 T  coincides with the $T_c$ measured from the onset of the drop of the resistance reported by Ni {\it et al.}~\cite{ni08}  in the same magnetic field. This is in contrast to the cuprates, especially the underdoped cuprates, where the Knight shift starts to decrease well above $T_c$, indicating the presence of a pseudogap.  At low temperatures in the superconducting state, we observe that the Knight shift decreases, almost linearly with temperature.

We express the total Knight shift as $K = K_s +K_{\mathrm{orb}} + K_{q} + K_{para}$.  The spin part of the shift is $K_s = A_{\mathrm{HF}}\chi_s$, where $\chi_s$ is the spin susceptibility, and $A_{\mathrm{HF}}$ is the hyperfine field.  Both the orbital, $K_\mathrm{orb}$, and the quadrupolar shift, $K_{q}$, are expected to be temperature independent.  Additional magnetic
shifts which we call $K_{para}$ could be associated with impurities,~\cite{muk09} and in principle, could be temperature dependent, although this is not seen in cuprates,~\cite{all09,bochen08} nor do we find any evidence for this in the present work.

In the normal state of Ba(Fe$_{0.93}$Co$_{0.07}$)$_2$As$_2$ below 170 K, we find the Knight shift can be described by an activated process.  However, there  is a substantial range at low temperatures where the Knight shift is essentially temperature independent, Fig.~\ref{fig2}. We use an Arrhenius form, $K(T) = A+ B\exp(-\Delta_{s}/T)$, where we find $\Delta_{s} = 394\pm16$ K, $A= 0.252\%$, and $B = 0.179$ with the fit shown by the (red) curve in the figure.  At high temperatures, 90 -- 170 K, the 
total Knight shift is proportional to the bulk susceptibility,~\cite{ni08} giving a hyperfine field of $A_{\mathrm{HF}}=14$ kOe/$\mu_B$, comparable to the value of 18.8 kOe/$\mu_B$ 
reported by Kitagawa {\it et al.}~\cite{kit08} Therefore the thermally activated process 
modifies the local density of states and directly affects the spin susceptibility. 
However, neither the origin of this high-temperature behavior, nor its relation to superconductivity is known.  The first observations of this effect, along with a similar analysis, was performed by Ning {\it et al.}~\cite{imai08} giving $\Delta_{s} = 520$ K for their 8 \% Co crystal of Ba(Fe$_{0.92}$Co$_{0.08}$)$_2$As$_2$.

In the superconducting state, we found that the total Knight shift 
gradually 
decreases on cooling from 12 K to 2 K in 11.6 T, figures 2 and 3. 
Similar behavior is consistently observed for higher magnetic fields up to 20 T, Fig.~4.  

If one takes the view that this is a linear temperature dependence, 
then such behavior can be associated 
with gap nodes at the Fermi surface, similar to cuprate superconductors with
$d$-wave symmetry.~\cite{takigawa91,hardy93,zheng02} 
The Knight shift $K_s(T)$ is proportional to the susceptibility  
\begin{equation}
\frac{\chi(H)}{\chi_n} = \int dE \, N(E) \frac{f(E-\mu H) - f(E+\mu H)}{2\mu H}
\label{eq:chi}
\end{equation}
where $\chi_n = 2 \mu^2 N_f$ - normal state susceptibility, 
$f(E)$  is the Fermi-Dirac distribution function, and $N(E)$ is the density
of states. 
For temperatures $T/T_c \gtrsim \mu H/2 T_c$ 
$[f(E-\mu H) - f(E+\mu H)]/[2\mu H] \approx - \partial f(E)/\partial E$ 
and for nodal quasiparticles $N(E) = E/\Delta_0$, resulting in 
a linear temperature dependence $\chi/\chi_n \sim T/T_c \times const\; O(1)$.  
This is a much steeper rise than what we observe.
Thus, we can exclude the possibility of nodes simultaneously present 
on both hole and electron sheets. 
However, nodes might be present on one of the pockets, 
presumably electronic, which in combination with the fully gapped 
hole pocket would give a slower $T$-dependence. 
Another possibility is that the observed temperature dependence is 
higher power than linear, and the electronic superconducting gap in 
Ba(Fe$_{0.93}$Co$_{0.07}$)$_2$As$_2$ is either anisotropic with weak ``accidental'' nodes, or has only a minimum 
with typical energy scale $\Delta_{min} < \mu H$, or even isotropically gapped, 
with strong pairbreaking due to impurities.~\cite{kim_can10,anton10} 
In this regard, the existence of impurities affecting the local magnetic fields 
is evident in the temperature dependence of
the NMR linewidth in the normal state, Fig.~5.

\begin{figure}[!ht]
\centerline{\includegraphics[width=0.49\textwidth]{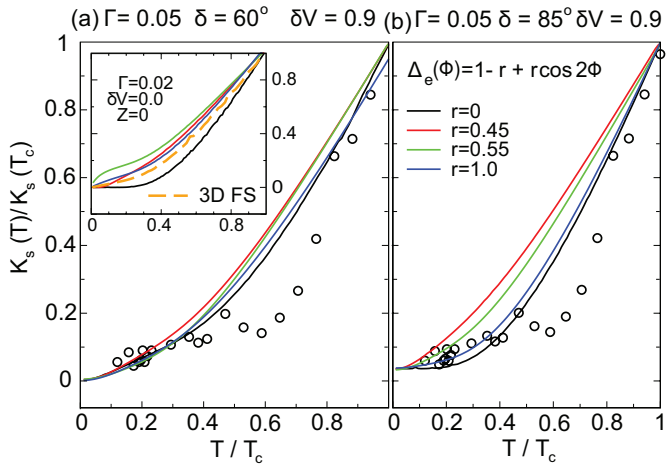}}
\caption{
Normalized susceptibility of Ba(Fe$_{0.93}$Co$_{0.07}$)$_2$As$_2$. 
The theoretical curves (solid lines) are shown for several 
possible gap structures 
on the electronic pocket, given by the form 
$\Delta_e(\phi) = \Delta_2(1-r + r\cos2\phi)$, for intermediate (panel a) and 
strong scattering (panel b). 
To compare these results with the experiment we took $Z = \mu_B H / T_{c0} = 0.3$ ($H=11$ T). 
Inset in panel (a) shows results for the clean case and at low fields ($Z=0$) - note 
that aside from the low-$T$ region all anisotropic gaps, $r=0.45,\,0.55,\,1$, 
and a model with more realistic 3D Fermi surface, 
behave in approximately the same way. 
In high field and 
for weak impurity scattering (a), all gap models are consistent with experiment at 
low $T$. 
In the unitary limit (b), the best fit is given by either combination of 
two isotropic gaps 
$\Delta_h(\phi) = -\Delta_e(\phi) = \Delta$ ($r=0$), 
or, surprisingly, by an isotropic hole 
$\Delta_h(\phi) = \Delta_1$ 
and d-wave like electron gaps
$\Delta_e(\phi) = \Delta_2 \cos 2\phi$ ($r=1$). It is possible that the dip in
the data, $ 0.5 < T/T_c < 0.8$, is a consequence of changes in the NMR
lineshape owing to vortex , section IV.
}
\label{fig3}
\end{figure}

Comparison of our Knight shift measurements for $H=11.6$ T with the above model and 
different gap structures are shown in Fig.~\ref{fig3}. 
The spin susceptibility is calculated according to Eq.~\ref{eq:chi}, 
where the shift of the energy levels in magnetic field is 
given by parameter $Z = \mu_B H/T_{c0} \sim 0.3$ ($H=11$ T), where 
$\mu_B$ is the Bohr magneton, and $T_{c0}=25$ K is an estimate of the zero-field 
transition temperature in the absense of impurities. 
The density of states is computed in the two-band model used 
previously and described in references 30 and 46.  
The gap on the hole FS is isotropic, $\Delta_h(\phi)=\Delta_1$, 
while on the electron FS it is a function of the angle 
$\Delta_e(\phi) = \Delta_2(1-r + r\cos2\phi)$, with $r=0$ being the isotropic 
gap, and $r=1$ gives four equally spaced nodes. 
The impurity scattering is given by the concentration parameter 
$\gamma = (n_{imp}/\pi N_f)/2\pi T_{c0}$, 
{\it interband} potential fraction ($\delta V= V_{12}/V_{11}$), 
and an {\it intraband} scattering phase shift ($\tan\delta= \pi N_f V_{11}$). 
The zero-temperature shift $K_s(T=0)$ is not well-defined experimentally and is left as a free parameter. 
Below $10\, T$ the temperature dependence is stronger and it is plausible that 
this can be associated with diamagnetic screening currents or vortex supercurrents which do not cancel as effectively in the intervortex region at low fields. 
At high temperature ($0.5< T/T_{c} <0.8$), 
the discrepancy between our data and the model is likely due to
thermal fluctuations of vortices that introduce a temperature dependence to the
NMR lineshape in a region where vortices are being pinned.   It is just in this
region of temperature that slow field fluctuations from vortices are manifest
in the linewidth, Fig.~5, and in spin-spin relaxation, discussed in the next section, Fig.~6.  We observe a systematic trend in the data of increasing Knight shift, $K(0)$, with increasing field, Fig.~4, which might be accounted for by considering the vortex core contributions to the density of states.

From Fig.~3, in the temperature range $T/T_c<0.5$ where we can compare 
the Knight shift data with the theoretical model, 
we find that more or less all electronic gap structures 
that we have considered in the $A_{1g}$ 
symmetry class can fit the data. 
One might say that the intermediate cases of anisotropic gaps with either 
minima $r=0.45$ or close nodes $r=0.55$ are less suitable, especially in the 
strong scattering case, leaving surprisingly 
a possibility of either an isotropically gapped electronic FS or 
one having a gap $\sim \cos2\phi$. 
A more reliable identification of the angular structure could be 
possible in lower magnetic fields ($Z=0$ inset in Fig.3(a)), 
but the analysis would be 
quite complicated due to the effects of vortices and diamagnetic screening currents.  
Different anisotropy parameters $r$ in this model give 
a reasonable coverage of the phase space of 
possibilities for the spectrum of quasiparticles.  
We checked that with a more realistic three dimensional 
Fermi surface and the gap suggested in Ref.~\onlinecite{mazin:2010_122nodes} 
results do not change much (dashed line in the inset of Fig.3(a) that 
appear exactly between the isotropically gapped $r=0$ and purely nodal $r=1$ 
cases).

The linewidth of the central transition, Fig.~\ref{fig5}, full-width-at-half-maximum (FWHM), was determined by fitting the spectra to a gaussian function. The fit deviates slightly near the wings but at low temperatures the lineshape appears to be symmetric.  The  linewidth increases with decreasing temperature in the normal state.  Below $T_c$, there is a decrease, followed by an upturn at $\sim$ 11 K in 11.6 T. These features closely parallel all of the aspects of the NMR linewidth previously reported by Chen {\it et al.}~\cite{bochen07,bochen08} for $^{17}$O NMR on high quality single crystals of Bi$_{2}$Sr$_{2}$CaCu$_{2}$O$_{8+\delta}$ (BSCCO). For cuprate superconductors\cite{all09,bochen08} this similar  behavior in the normal state was identified with impurities.  Its classic signature  is a Curie or Curie-Weiss temperature dependence to the linewidth and a constant Knight shift.  The basic argument is that impurities introduce strong polarization of the nearby 
conduction electrons which produce a spatially oscillating spin density that couples through the 
hyperfine interaction to the nuclei (RKKY-interaction).  Consequently, the average
local field  is not significantly perturbed, so the Knight shift is unaffected; but, a large distribution of Knight shifts gives a significant broadening of the NMR spectrum.   The polarization of paramagnetic impurities is typically quite temperature dependent (Curie-like), and is the likely origin for the temperature dependence in the linewidth below $T \lesssim 80$ K.  At higher temperatures there is a possible additional temperature dependence from the activated process that is evident in the Knight shift shown in Fig.~2.

The success we have had with the above model in understanding the $^{17}$O NMR linewidth in BSCCO,~\cite{bochen08} serves as a guide for our interpretation of the impurity contributions to the $^{75}$As NMR linewidth in Ba(Fe$_{0.93}$Co$_{0.07}$)$_2$As$_2$. We focus on the range of
temperature where the Knight shift is mainly temperature independent, $T_c < T \lesssim 80$ K.
We fit the linewidth to the following
phenomenological relation,

\begin{equation}
\Delta\nu = \Delta\nu_0\left (1-a\exp(-\Delta_{s}/k_{B}T)\right )+\frac{CH_0}{T} K_s(T).
\label{eq4}
\end{equation}

\noindent Since there may also be a
distribution of Knight shifts with a thermally-activated origin, we allow for
this contribution with the additional term, $a\exp(-\Delta_{s}/k_{B}T) $, which goes beyond the 
impurity model used for BSCCO.\cite{bochen08} Here, $\Delta_{s} = 394$ K,  obtained from a fit to $K(T)$ in the 
normal state with $K_{orb} = 0.229\%$ with parameters, $\Delta\nu_0, C$, and $a$.   Overall, the fit is reasonable, capturing the essential
characteristics of the normal state behavior of the linewidth with a reduction in the superconducting state as the impurity local field distribution becomes gapped following the temperature dependence of the Knight shift.  Although impurity effects can account for the temperature dependent linewidth, we cannot comment on what type of impurity might be responsible. 

The upturn of the linewidth at $T_m \sim 11$ K in 11.6 T in the superconducting state is the mark of  vortex freezing. It occurs at $T_m/T_c = 0.61$ in $H = 11.6$ T, and indicates that there is a substantial temperature  range below $T_c$ where fluctuating vortices are in a liquid-like phase. We can compare this with
YBa$_{2}$Cu$_{3}$O$_{7-y}$ (YBCO)~\cite{rey97} and overdoped BSCCO~\cite{bochen07} where vortex melting for the same magnetic field occurs  at $T/T_c = 0.87$ and $0.21$ respectively. Calculations of the melting transition in high magnetic fields for pnictide superconductors have been performed by Murray and Tesanovic~\cite{mur10}.  However, a true, thermodynamic, vortex-liquid phase in Ba(Fe$_{0.93}$Co$_{0.07}$)$_2$As$_2$ would be surprising since both the mass anisotropy\cite{yam09} and $T_c$ are small ; for example, as compared to cuprate superconductors.  It was argued by Yamamoto {\it et al.}~\cite{yam09} that a confluence of weak thermal fluctuations and a very broad distribution of pinning forces gives rise to an irreversibility field significantly less than the upper critical field in crystals of Ba(Fe$_{0.9}$Co$_{0.1}$)$_2$As$_2$.  Generally the NMR spectrum is sensitive to field fluctuations on time scales slower than the Larmor precession period which is tens of nanoseconds in our case.   Consequently, the sharp variation in the linewidth that we observe at $T_m$ suggests that vortex dynamics are slower than this for $T<T_m$.  To investigate this further,  we have performed spin-spin relaxation experiments discussed in the next section.

Usually, the vortex contribution to the NMR or $\mu$SR linewidth\cite{rey97,son00} is interpreted in terms of the penetration depth.  In Ba(Fe$_{0.93}$Co$_{0.07}$)$_2$As$_2$ the linewidth broadening from the vortex solid state can be seen in Fig.~\ref{fig5}  as a difference between data (represented by a black dashed line) and our model for the vortex free linewidth (solid blue curve) discussed above. The vortex component of our linewidth at the lowest temperatures, $\sim 7.5$ kHz ($\sim 10$ gauss) is much less than predicted by Ginzburg-Landau theory~\cite{bra03} which is 33 kHz for a penetration depth from $\mu$SR of $\lambda_{ab} = 217$ nm,~\cite{wil10}  or  15 kHz for the value, $325 \pm 50$ nm, from scanning SQUID and magnetic force measurements.\cite{lua10}  The most likely reason for this discrepancy is c-axis vortex disorder discussed by Brandt.~\cite{bra91} For example, at very low magnetic fields, $H\lesssim 1$ T, this plane-to-plane disorder is well-established in BSCCO from $\mu$SR and NMR measurements~\cite{inu93,bochen07,mou10a,mou10b} and can be expected in Ba(Fe$_{0.93}$Co$_{0.07}$)$_2$As$_2$ since in-plane disorder and strong pinning has already been established by scanning probe measurements~\cite{lua10,yin09} and small angle neutron scattering.~\cite{esk09}

\section{Spin-spin Relaxation and Vortex Dynamics}

To explore possible effects of vortex dynamics, already indicated in our NMR linewidth measurements, we have investigated spin-spin relaxation. There are two classes of spin relaxation experiments that we perform.   The results measured by Hahn echo methods are presented in Fig.~6 and 7, and  by the CPMG pulse sequence in Fig.~6 and 8.   The nuclear spin dephasing times, $T_{2H}$ and $T_{2CPMG}$, from these experiments can be viewed semi-classically as a measure of field fluctuations along the direction of the applied magnetic field.  The dephasing rate is a summation of  contributions from spin-lattice relaxation (Redfield contribution), local fluctuations in the external magnetic field, and fluctuations in the internal fields such as those from the nuclear dipole-dipole interaction, or from vortex dynamics.  The Hahn echo method is generally useful for homogeneous relaxation (all nuclei are equivalent) in a steady external field.  However, if the local field fluctuates inhomogeneously, or there is nuclear spin diffusion in a spatially inhomogeneous steady field,\cite{gen75,yu80} then the  CPMG method can be helpful in identifying the  time scale of the fluctuations through variation of the spin refocusing time, $\tau$ (see the experimental methods section).~\cite{sigmund01}

From our relaxation measurement data on Ba(Fe$_{0.93}$Co$_{0.07}$)$_2$As$_2$ shown in Fig.~6, it is immediately apparent that $T_{2CPMG}$ is significantly longer than $T_{2H}$ and that in the normal state, both are much longer than the temperature independent dipolar contribution,  $0.7$ ms, calculated from the rigid lattice limit, shown as a red dashed line.  For all CPMG experiments  shown here $2\tau$ was set to be relatively short, $250 \,\,\mu$s to  $700 \,\,\mu$s.  We infer that significant dephasing takes place on time scales longer than $700 \,\,\mu$s and, since this is in the normal state, the process responsible has nothing to do with vortices.  Rather it is an indication of the presence of slow field fluctuations from an unidentified source.   In fact, by varying the refocusing time, $2\tau$, we have determined that this dephasing is effectively eliminated in the CPMG results which we report here.  We defer  our discussion of the mechanism for the Hahn echo relaxation  to later in this section. 

The temperature dependence of $T_{2CPMG}$ in the normal state can be understood from the temperature dependence of spin-lattice relaxation, through the Redfield contribution.~\cite{redfield67}  This behavior can be expressed phenomenologically as, $T_{2CPMG}^{-1}=T_{2}^{-1} + (\alpha T_{1})^{-1}$.~\cite{butterworth} Taking $T_{1}$ from Ning {\it et al.}~\cite{imai08} we find $\alpha$ = 5.7 and $T_{2}=12$ ms, where $T_2$ is the spin-spin relaxation time after correction for this contribution. Since $T_1$  is long enough below $T_c$ ($>$ 150 ms) we may take $T_{2CPMG}$ to be equal to $T_2$ in the superconducting state.  Nonetheless, we find that  there is a remarkable dip  in both $T_{2CPMG}$ and in $T_{2H}$ at $T_m$, Fig.~6 to 8, which we identify with thermal fluctuations of vortices.   This follows from the fact that the temperature of the minimum in $T_2$ coincides with the vortex freezing temperature apparent in the NMR linewidth, presented in Fig.~5 for 11.6 T and that this dip at $T_{m}$ appears just below the upper critical field, Fig.~ 9.  

Vortex dynamics  contribute  similarly in these two types of spin-spin relaxation experiments although the contribution to the Hahn echo can be an order of magnitude larger than for the CPMG, especially at higher magnetic fields.  This means that the principal fluctuation time scale must be  of order  $T_{2H}$, but longer than the refocusing time, $2\tau = 700 \,\,\mu$s, in the CPMG measurement.  

In previous work on $^{17}$O NMR in YBCO it was found that the spin-spin relaxation rate, $1/T_2$, had a peak below $T_c$.~\cite{bachman98,recchia97,suh93,curro00,kramer99} Kr{\"a}mer {\it et al.}~\cite{kramer99}  proposed a dynamical charge-density-wave state  in YBa$_2$Cu$_3$O$_{7-\delta}$ coupled to the nucleus through the quadrupolar interaction.  However,  Bachman {\it et al.}~\cite{bachman98} found that this peak was an artifice of vortex dynamics which become slow enough in the vortex solid to be observable in $T_2$.  They found a lorentzian component to the spectral density for field fluctuations that abruptly onsets at the irreversibility temperature.  The fact that the onset temperature was magnetic field dependent indicated that the phenomenon was not related to a charge density wave.   This argument holds equally well in the present case for Ba(Fe$_{0.93}$Co$_{0.07}$)$_2$As$_2$.  Here, as for the cuprates, we associate the temperature of the maximum spin-spin relaxation rate (the dip in the relaxation time) with the temperature where the vortex dynamics rapidly slow down, {\it i.e.} the irreversibility temperature.  For comparison, in Fig.~9 we plot, along with our data,  the irreversibility curve  from  Prozorov {\it et al.}~\cite{prozorov08} obtained at lower magnetic fields for similar crystals from the Ames Laboratory.  Our data includes the temperatures of anomalies from the NMR linewidth (spectrum) and $T_2$ measurements we have discussed previously.  The CPMG sequence was used to obtain $T_2$ at the higher magnetic fields,  20 T to 28 T because the Hahn echo experiment is adversely affected by magnetic field fluctuations endemic to high field resistive magnets.~\cite{sigmund01}   We also compare our measurements of the superconducting transition  with those of Ni {\it et al.}~\cite{ni08,prozorov08}  providing a consistent vortex phase diagram for Ba(Fe$_{0.93}$Co$_{0.07}$)$_2$As$_2$.

Now we return to a discussion of the large $T_2$ we have observed, as compared with the expected temperature independent upper bound established by the  nuclear dipole-dipole interaction, the red dashed line in Fig.~6.  In order for $T_2$ to be longer than the dipole limit there must be some form of coherent averaging which in liquids is the well-known phenomenon of motional averaging.  In metals there is a similar mechanism, a type of spin-motional averaging, which is attributable to a strong indirect interaction between  adjacent nuclei, $^{75}$As in our case.~\cite{aw53,yu80}  For this process to work the coupled nuclei must be resonant, that is to say they must have the same Larmor frequency to within the small band given by the bare nuclear dipole-dipole interaction.  This requirement is imposed by the necessity to conserve energy in the spin `flip-flop' process for spin-spin relaxation.  The indirect interaction is a coupling between nuclei, mediated by the conduction electrons, involving an RKKY type polarization and the hyperfine interaction.  For metallic elements with only one spin isotope this leads to averaging of the dipolar local fields, called exchange narrowing, and can substantially increase $T_2$, a factor of 5.6 in the case of platinum.~\cite{yu80,Froidevaux68}  For elements with more than one spin isotope the nuclei are non-resonant and the interaction leads to $T_2$ shorter than the dipolar limit, as is the case for thallium.  From the theory of exchange narrowing~\cite{aw53} we have,

\begin{equation}
 1/T_2 \simeq\omega_{dipole} ^2 /\omega_{e}
\end{equation}

\noindent where $\omega_{e}$ is the frequency of the exchange interaction and $\omega_{dipole}$ is the bare dipolar interaction between nuclei.  From this relation we find, $\omega_{e}\simeq24$ kHz,  which is larger than the 4 kHz,~\cite{Froidevaux68} observed for metallic platinum, but not unreasonably large given that $^{75}$As is 100 \% naturally abundant as compared to 33.8 \% for $^{195}$Pt.   Although this explanation can account for the longer $T_{2}$ than expected from the dipolar interaction, it does not explain the shorter phase coherence time found from Hahn echo measurements as compared with the CPMG sequence.

There are two possibilities for the source of low frequency field fluctuations that could give rise to the difference between Hahn and CPMG results.   In the first, we consider fluctuating magnetic fields, either from impurities or residual effects of the suppressed antiferromagnetism inherent to the BaFe$_2$As$_2$ system.  However, it is hard to reconcile this explanation with our observation that $T_{2H}$ is independent of  magnetic field and temperature, as is evident in Fig.~7 after allowing for the vortex contribution.  One might expect that thermally driven, magnetic field fluctuations would be both temperature and magnetic field dependent over the wide range we have covered.  A second possibility is that the rather large indirect interaction leads to nuclear spin diffusion that produces dephasing in a locally inhomogeneous, static, magnetic field.  We have estimated the diffusion coefficient from the indirect interaction\cite{red68,yu80} to be $D_s \approx 10^{-12}$ cm$^2$/s which, in order to explain our Hahn echo results, would require an average magnetic field gradient of $\approx$ 1.5 $\times$ 10$^7$ G/cm.  Associating a magnetic field inhomogeneity of this size with residual magnetism in the sample could provide the temperature independent mechanism that would be required to account for the Hahn echo results.  The length scale for the distribution of such magnetism is constrained by the linewidth to be less than $\approx 20$ nm.  Although it may seem that the latter explanation is the more likely, further work will be necessary to confirm or disprove either of these suggestions.

\section{Conclusion}

We have studied $^{75}$As NMR spectra and spin-spin relaxation on a single
crystal of optimally doped Ba(Fe$_{0.93}$Co$_{0.07}$)$_2$As$_2$ from room
temperature to well below the superconducting transition temperature in
magnetic fields as high as 28 T.  The temperature dependence of the Knight
shift is compatible with a general, two-band $A_{1g}$ symmetry scenario for the superconducting 
states.  The details, nodes or full gap, cannot be determined from the present 
work.  Comparison with an elementary 2D model might indicate that the 
order parameter in this compound has nodes but they have relatively 
weak weight compared to the fully gapped portions of the Fermi surface.   
We have identified a sharp signature of irreversibility in the temperature
dependence of our measured spin-spin relaxation from which we have established
a vortex phase diagram  up to high magnetic fields.  We have found that there
is a very strong indirect exchange interaction in this compound providing a
possible explanation for our observations of spin-spin relaxation attributed to
nuclear spin diffusion in static magnetic field gradients.\\

\section{Acknowledgments}

We thank C.A. Collett, W.J. Gannon, J. Li, and J. Pollanen for  critical reading of the manuscript and acknowledge support from the Department of Energy, Basic Energy  Sciences
under Contracts No. DE-FG02-05ER46248 (Northwestern University)  
and No. DE-AC02-07CH11358 (Ames Laboratory). Work at high magnetic field was performed
at the National High Magnetic  Field Laboratory with support from the National
Science Foundation and the State of Florida.  A.B.V. acknowledges support from NSF grant DMR-0954342.

\newpage

\section{Figure Captions}

\begin{figure}[!ht]
\centerline{\includegraphics[width=0.4\textwidth]{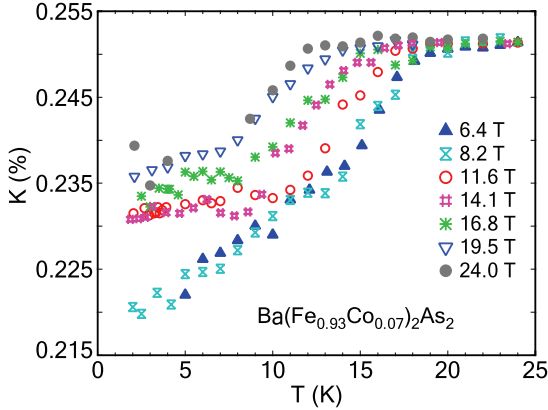}}
\caption {Knight shifts at various magnetic fields.   The low field values have a strong temperature dependence at low temperatures.  This is ascribed to vortex supercurrent contributions to the line shape which are expected to become negligible at high fields.  The weak temperature dependence of the Knight shifts for $H> 11$ T at the lowest temperatures, should be independent of vortex supercurrents. An increase in Knight shift, $K(0)$, with increasing magnetic field is observed.}
\label{fig4}
\end{figure} 
\begin{figure}[!ht]
\centerline{\includegraphics[width=0.4\textwidth]{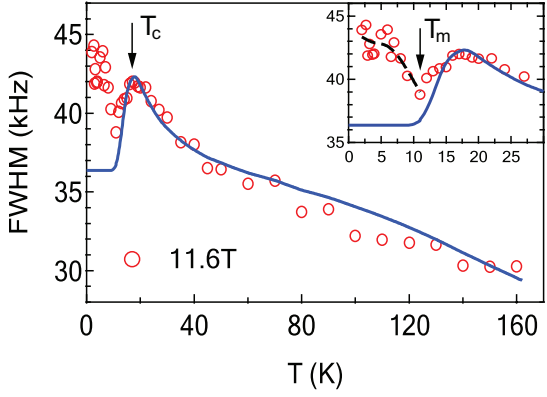}}
\caption { Linewidth of the $^{75}$As central 
transition at $H = 11.6$ T. The upturn of the linewidth at $T_m \sim$ 11 K, indicates freezing of the vortex liquid\cite{bochen07} (the inset shows an expanded view of this transition). The solid blue curve, down to $T_m$, is a fit to the model described in the text. Below $T_m$ the solid blue curve is a calculation from this model, absent any vortex contributions to the linewidth.  The dashed black line is a guide-to-the-eye representing the data.  The difference at low temperatures between the solid and dashed curves in the inset is  the vortex contribution to the linewidth which we find to be $\approx 7.5$ kHz.  }
\label{fig5}
\end{figure}

\begin{figure}[!ht]
\centerline{\includegraphics[width=0.4\textwidth]{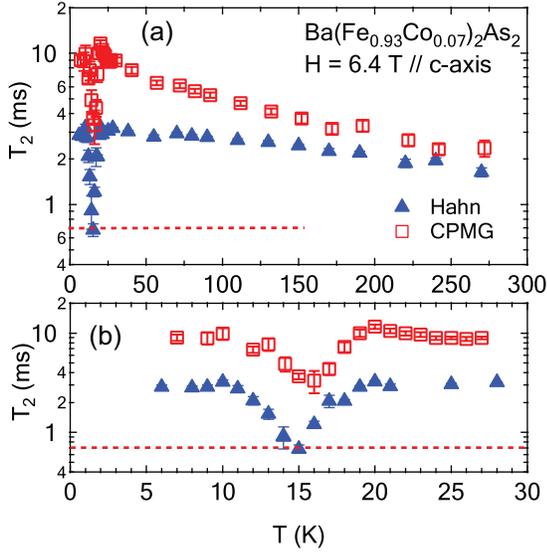}}
\caption {Spin-spin relaxation times at $H=6.4$ T.  Measurements of spin-spin relaxation times with Hahn echo and CPMG methods are shown for comparison. A superconducting magnet was used to rule out $T_2$ reduction due to field fluctuation. a) The slow increase with decreasing temperature down to $T_c$ corresponds to reduction in the Redfield contribution, {\it i.e.} from spin-lattice relaxation.  b) An expanded view at low temperature below $T_c$, shows that there is a  decrease in both relaxation times.  We associate this with vortex dynamics providing a maximum contribution to the rates at $T_m$, taken from Fig.~5. The red dashed line indicates $T_2$ in the dipolar limit, 0.7 ms.}
\label{fig6}
\end{figure}

\begin{figure}[!ht]
\centerline{\includegraphics[width=0.4\textwidth]{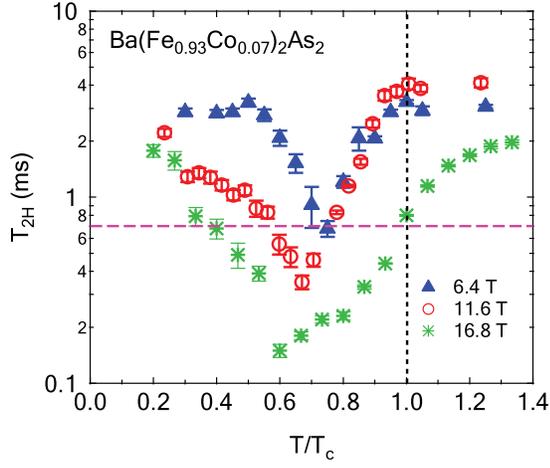}}
\caption {Hahn echo relaxation times, $T_{2H}$, for various magnetic fields from a superconducting magnet. There is a clearly defined temperature, $T_m$, for a minimum in $T_2$ at each magnetic field which we associate with vortex freezing. The minimum $T_2$ becomes even smaller than the $T_2$ from the dipolar limit (0.7 ms) when the magnetic field increases.  Additionally, $T_{2H}$ shows a decrease in the normal state just above $T_c$ at $H=16.8$ T.}  
\label{fig7}
\end{figure}

\begin{figure}[!ht]
\centerline{\includegraphics[width=0.4\textwidth]{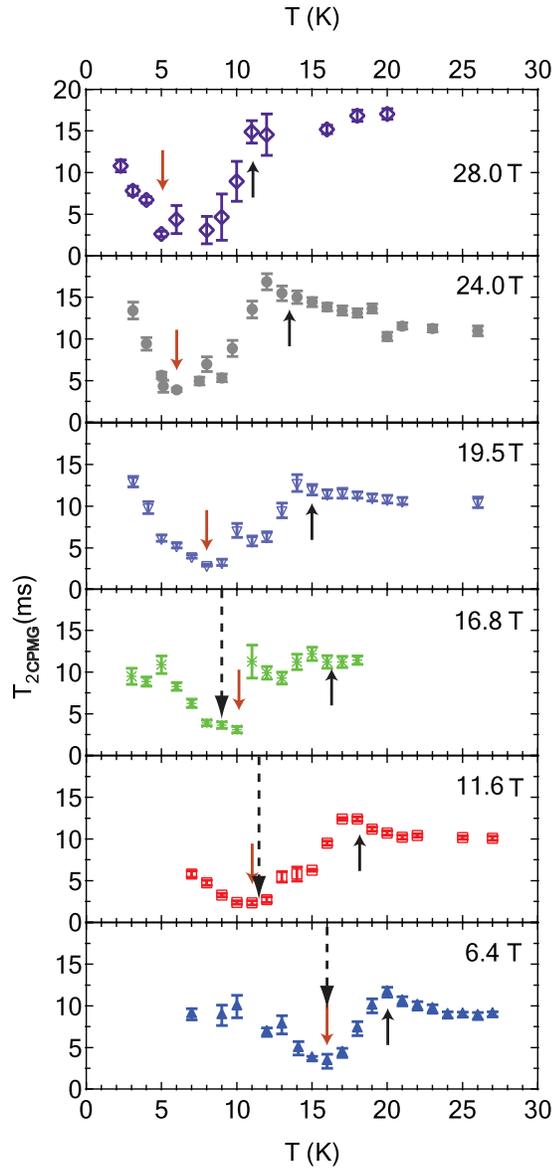}}
\caption {CPMG relaxation times, $T_{2CPMG}$, for various magnetic fields.   The transition temperature\cite{ni08} (black upward arrow) and the vortex freezing  temperature (red downward arrow) correspond to abrupt changes in relaxation. The dashed black arrows indicate the minimum relaxation time from the Hahn echo method.}
\label{fig8}
\end{figure}

\begin{figure}[!ht]
\centerline{\includegraphics[width=0.4\textwidth]{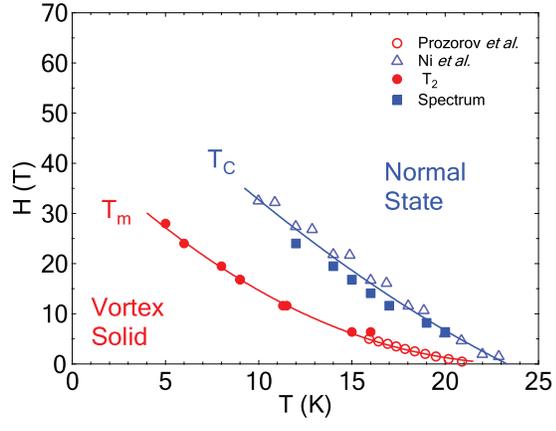}}
\caption {Vortex phase diagram.  The vortex freezing temperature deduced from the minimum in the spin-spin relaxation time and from the minimum in the linewidth is shown as a function of magnetic field along with the vortex irreversibility temperature from Prozorov {\it et al.}~\cite{prozorov08}  The superconducting transition temperature inferred from the NMR spectrum linewidth and Knight shift are compared with the report from Ni {\it et al.}~\cite{ni08}}
\label{fig9}
\end{figure}

\end{document}